\documentclass[a4paper]{article}
\usepackage{multirow}
\usepackage{graphicx}

\usepackage{INTERSPEECH2020}

\title{Acoustic Word Embedding System for Code-Switching Query-by-example Spoken Term Detection}
\name{Murong Ma$^{1,4}$, Haiwei Wu$^{1,5}$, Xuyang Wang$^{3}$, Lin Yang$^{3}$, Junjie Wang$^{3}$ and Ming Li$^{1,2}$}
\address{
$^{1}$Data Science Research Center, Duke Kunshan University, Kunshan, China \\
$^{2}$School of Computer Science, Wuhan University, China\\
$^{3}$AI Lab of Lenovo Research, Beijing, China\\
$^{4}$Computer Science and Technology, Beijing University of Post and Telecommunications, China\\
$^{5}$School of Electronics and Information Technology, Sun Yat-sen University, Guangzhou, China}
\email{ming.li369@dukekunshan.edu.cn}

\begin{document}

\maketitle
\begin{abstract}
In this paper, we propose a deep convolutional neural network-based acoustic word embedding system on code-switching query by example spoken term detection. Different from previous configurations, we combine audio data in two languages for training instead of only using one single language. We transform the acoustic features of keyword templates and searching content to fixed-dimensional vectors and calculate the distances between keyword segments and searching content segments obtained in a sliding manner. An auxiliary variability-invariant loss is also applied to training data within the same word but different speakers. This strategy is used to prevent the extractor from encoding undesired speaker- or accent-related information into the acoustic word embeddings. Experimental results show that our proposed system produces promising searching results in the code-switching test scenario. With the increased number of templates and the employment of variability-invariant loss, the searching performance is further enhanced.

\end{abstract}

\section{Introduction}

Spoken term detection (STD) \cite{Mandal2014Recent, Park2008Unsupervised} is a technique to detect specific words in streaming audio or audio files. With the development of Internet media and smart devices, the demand for searching keywords in audio signal and voice control increased rapidly. 

Query by example (QbE) is a special case of the STD problem, whose task is to find out the occurrences of a keyword given its audio samples. A typical solution for this task is applying Dynamic Time Warping (DTW) \cite{muller2007dynamic} or its variants on frame-level features extracted from keyword templates and searching content \cite{Hou2014TheNQ}. Both supervised \cite{Rodriguez2014High,Leung2016high} and unsupervised \cite{chen2016unsupervised,Glass2009Unsupervised} methods are explored to extract frame-level features by many researchers. Unsupervised features contain traditional acoustic features like filter-bank energy (Fbank) and Mel-Frequency Cepstrum Coefficients (MFCC) \cite{Yang2014Intrinsic}, as well as features obtained from unsupervised models like GMM by computing the posterior probabilities of the components \cite{Glass2009Unsupervised}. Supervised frame-level features include phonetic features extracted by a neural network like language-independent Bottleneck feature (BNF) and phone posterior probabilities \cite{Kramer1991Nonlinear, Hou2014TheNQ}. DTW and its variants, such as segmental DTW\cite{muller2007dynamic} and subsequence DTW \cite{muller2007dynamic,Anguera2013Memory,harry2018concatenate}, are then employed to find out the most matching feature sequences in the audio content with templates. 

In recent years, there has been increased interest in applying acoustic word embedding (AWE) methods to QbE-STD tasks \cite{Kamper2016Deep, Settle2016DiscriminativeAW, Settle2017Query, levin2017segmental}. Acoustic word embeddings are segment-level features extracted from the penultimate or final layer of word discriminate neural network. The network projects the features of audio segments to a fixed-dimensional vector space. Researchers have explored different network structures in this task, including convolutional neural networks (CNNs) \cite{Kamper2016Deep} and recurrent neural networks (RNNs) \cite{Settle2016DiscriminativeAW, Settle2017Query}, which show superior performance to traditional methods. After training, embeddings of the same keyword have smaller distances with each other, and embeddings of the different keywords have more considerable distances. Then, a sliding analysis window \cite{Yuan2018Learning} is taken to detect the occurrences of keywords. 

Most of the AWE systems focus on the single language scenario, which means that only one language is spoken in the audio. However, speakers may switch between several languages in real-life situations. Code-switching is a practice of alternating between two or more languages in the context of a single conversation, and it is a common phenomenon in many areas of the world, especially in second language education. However, to our knowledge, most studies focusing on the code-switching scenario are automatic speech recognition (ASR) tasks oriented and fewer studies on the QbE-STD task. This motivates us to explore the QbE-STD task from the code-switching testing data point of view.

In this paper, we propose a multi-language deep acoustic word embedding system with multiple templates. Besides the word discrimination loss, an auxiliary variability-invariant loss is also proposed to make the system generalize better on searching content and keywords spoken by different speakers. The word discrimination loss learns to encode word embeddings with labeled word audio, and the variability-variant loss aims to further decrease the distance between embeddings of the same keyword spoken by different speakers with or without accents. The similar method has also been used in fields such as speaker recognition \cite{danwei2020with-in}, speech recognition \cite{liang2018spoken} and and far-field speaker recognition\cite{Huang2019intel}. Our method selects audio data in English and Chinese for our training instead of using training data from a single language in other configurations. With data in two languages, We train a deep convolutional neural network for audio word discrimination. The trained model is used to extract embeddings for both keyword audio and searching content, and a sliding window accompanied by cosine distance computation is applied to detect the keywords. We also utilize the averaged template to reduce the within keyword variabilities. Recently, this idea is also applied in the QbE-STD system \cite{Yuan2019Verifying}.  

\begin{figure*}[th]
    \centering
    \includegraphics[scale=0.5]{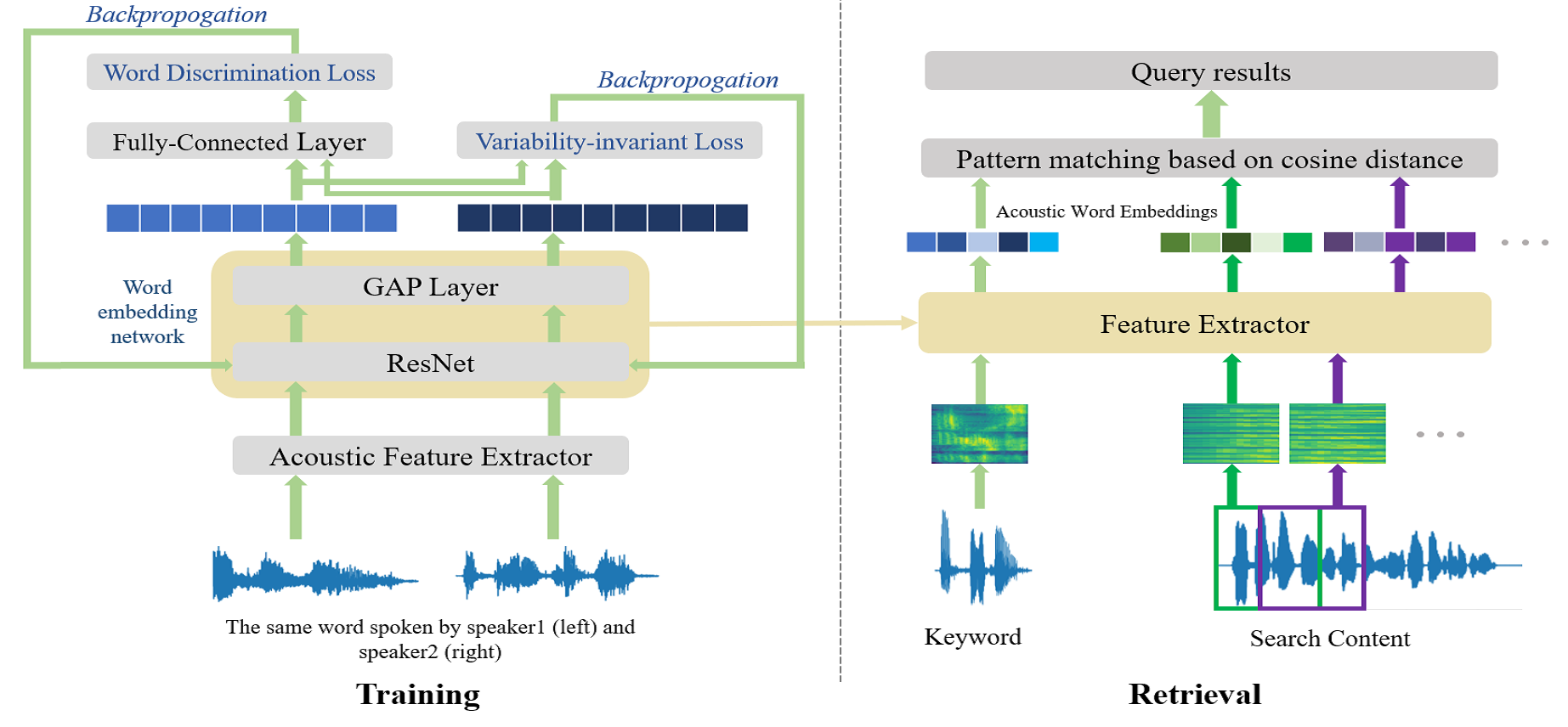}
    \caption{The pipline of the whole AWE system for the QbE-STD task}
    \label{fig:awe}
    \vspace{-0.5cm}
\end{figure*}
\section{Baseline systems}

% Generally speaking, there are two main steps in the traditional query by example (QbE) system, feature extraction, and template matching. The system performance mainly depends on the quality of input features, as well as the accuracy and time complexity of the matching algorithm. In previous works, various kinds of features have been applied to represent query examples and search content in an informative way, including acoustic features like Fbank, MFCC, and phonetic features like phone posteriors probabilities of GMM or DNN models. Among those, DNN phone posterior probabilities (PPP) achieve the state-of-the-art performance in traditional QbE methods. Dynamic time warping (DTW) and its variants are the most commonly used matching algorithms.
Generally speaking, there are two main steps in the traditional query by example (QbE) system, feature extraction, and template matching. In our baseline system, we utilize the DNN based phone posterior probabilities (PPP) as feature together with subsequence dynamic time warping (subsequence-DTW) algorithm for template matching.

\subsection{DNN phone posterior probabilities}

DNN based acoustic modeling is usually applied in automatic speech recognition (ASR) task and achieves state-of-the-art performance. In our study, we employ an advanced acoustic modeling method based on the time delay neural network (TDNN) to obtain the phone posterior probabilities (PPP) as the baseline system. To make the baseline system suitable in a code-switching scenario, we separately train a Chinese acoustic model and an English acoustic model for phonetic feature extraction using the Kaldi toolkit \cite{povey2011kaldi}. The final PPP feature sequences can be fetched from the output layer of the trained TDNN acoustic model.

\subsection{Template matching}
The matching algorithm and fusion method used in the baseline system will be introduced in this section.

\textsl{a) Subsequence-DTW.} Traditional DTW \cite{muller2007dynamic} requires the start and the end of two sequences must be strictly aligned. Instead, in our cases, we employ subsequence-DTW (S-DTW) \cite{Anguera2013Memory,harry2018concatenate} which allows us to find a subsequence within the search content that most optimally fits the spoken query. 

\textsl{b) Fusion method.} In this work, we employ multiple templates strategy in both baseline system and our proposed system with different fusion methods. Averaged template is widely used to cope with the variability caused by extraneous factors like noise and different speakers in the QbE-STD task. In the baseline system, A DTW-based fusion scheme is applied to obtain a more representative example \cite{chen2015QbE}. Specifically, in the first step, we randomly choose one from the prepared templates as the main template. And secondly, we apply the DTW algorithm to align each of the rest templates with the main template to get a warping path. Thirdly, we can calculate the overall average of all prepared templates for each aligned point with the warping paths for each keyword and obtain a final template representing the keyword.

% Ideally, we hope features encoded from the same word has minimal distance. However, in real life, variability exists because of , and other factors. In the baseline system, A DTW-based fusion scheme is applied to obtain a more representative example \cite{chen2015QbE}. Specifically, in the first step, we randomly choose one from the prepared templates as the main template. And secondly, we apply the DTW algorithm to align each of the rest templates with the main template to get a warping path. Thirdly, we can calculate the overall average of all prepared templates for each aligned point with the warping paths for each keyword and obtain a final template representing the keyword.

\section{Acoustic Word Embedding}

% Recently, methods of acoustic word embedding show excellent performance in the task of QbE-STD. AWE system maps the segments of spoken terms to a vector space where cosine distance or other distance functions can be introduced to evaluate the similarity of different audio words. This method also achieves excellent results in many other speech areas like speaker verification and paralinguistic attributes recognition \cite{Cai2018Exploring}. The auxiliary variability-invariant loss is shown to be effective in speaker verification under noisy environment\cite{danwei2020with-in}. In our work, we use a similar configuration to constrain the distance or distortion between embeddings of the same keyword spoken by different speakers. The network structure and the matching algorithm will be introduced in section 3.1. and 3.3.

\subsection{Network structure}

\begin{table}[b]
    \normalsize
    \centering
    \caption{ResNet structure. N/A: Not available}
    \label{resnetconfig}
    \resizebox{0.98\columnwidth}{!}{
        \renewcommand\arraystretch{1.3}
        \begin{tabular}{|c|c|c|c|c|}
            \hline
            Layer  & Output size & Downsample &  Channels &  Blocks  \\ \hline
            Conv1  & 16 $\times$ $\frac{L}{4}$  & False & 64  & -    \\ \hline
            Res1   & 16 $\times$ $\frac{L}{4}$  & False & 64  & 3    \\ \hline
            Res2   & 8 $\times$ $\frac{L}{8}$  & True & 128  & 4    \\ \hline
            Res3   & 4 $\times$ $\frac{L}{16}$  & True & 256  & 6   \\ \hline
            Res4   & 2 $\times$ $\frac{L}{32} $  & True & 512 & 3    \\ \hline
            GAP        & 512 &N/A&N/A&N/A \\ \hline    
            Output     &  number of words &N/A&N/A&N/A \\ \hline    
    \end{tabular}}
\end{table}

Our network is a combination of a convolutional neural network structure, a global average pooling layer, and a fully-connected layer in sequence. Log filter-bank energies (Fbank) are used as the input acoustic feature sequences. The CNN structure works as a local pattern extractor that maps the input feature sequences into a compressed high-level abstract tensor block with temporal order. We set up our deep CNN based on the popular residual neural network (ResNet) \cite{He2016Deep}. The corresponding parameters are described in Table \ref{resnetconfig}. 
% ResNet is widely used as a feature extractor and classifier in an end-to-end multi-layer fashion. It overcomes the degradation problem by realizing a shortcut connection, which makes it easy to optimize and improves the classifying accuracy. 

By forwarding the feature sequences through the deep CNN structure, the acoustic features can be transformed into a three-dimensional feature map, which still has one dimension related to time. Then the following global average pooling layer (GAP) acts as an aggregator over the entire sequence by computing the global mean feature values over the time and frequency axes. The output representation is then fed into the following fully-connected (FC) layer. Each unit in the output layer is labeled with an individual word in training data. Cross-entropy loss and auxiliary variability-invariant loss are employed to optimize the system, and the final acoustic embedding can be fetched from the output of the GAP layer. The whole procedure of AWE system for the QbE-STD task is depicted in Figure~\ref{fig:awe}.

% \begin{figure}
%     \centering
%     \includegraphics[scale=0.4]{fig/oneblock.pdf}
%     \caption{Network structures of block softmax (left) and one softmax (right) used in multilingual AWE system}
%     \label{fig:blocksoftmax}
%     \vspace{-0.5cm}
% \end{figure}

% In addition to the traditional one softmax layer, we also employ block softmax layer, which has proved effective in multilingual  BNF extraction \cite{Kramer1991Nonlinear, Hou2014TheNQ}. The major difference of our study is that we employ block softmax on segment-level input targeted with words instead of frame-level input targeted with phonemes. The main idea of block softmax is to lead language-dependent information into the feature map by dividing the output layer into multiple blocks according to the language by using structure illustrates in Figure~\ref{fig:blocksoftmax}. Each block of the output layer corresponds to an individual language and is activated only if the input data is from the associated language. This mechanism can be implemented with an interval-based softmax function 

In addition to the traditional one softmax layer, we also employ block softmax layer, which has proved effective in multilingual  BNF extraction \cite{Kramer1991Nonlinear, Hou2014TheNQ}. The major difference of our study is that we employ block softmax on segment-level input targeted with words instead of frame-level input targeted with phonemes. The main idea of block softmax is to lead language-dependent information into the feature map by dividing the output layer into multiple blocks according to the language. Each block of the output layer corresponds to an individual language and is activated only if the input data is from the associated language. This mechanism can be implemented with an interval-based softmax function 

\begin{equation}\label{equ:block softmax}
    y_i=\frac{exp(a_i)}{\sum_{j=n_{l,b}}^{n_{l,e}}exp(a_j)}.
\end{equation}
$y_i$ denotes the posterior of the i-th output; $a_i$ represents the i-th activation value and $n_{l,b}$ is the beginning index of the \textsl{l}-th language while $n_{l,e}$ is the ending index.

\subsection{Variability-invariant Loss}

Recently, variability-invariant loss has been employed in speaker recognition \cite{danwei2020with-in}, speech recognition \cite{liang2018spoken} and far-field speaker recognition \cite{Huang2019intel}. For QbE-STD tasks, the person who speaks search content and keyword are usually different. Ideally, the embeddings of the same keyword spoken by different speakers should be identical to each other. However, the extractor usually encodes speaker-related information as a part of the word representation. To make it concentrate more on word discrimination, we use the variability-invariant loss for each word during the training phase. For each instance $I_{w,p_1}$ with word label $w$ and speaker label $p_1$ forwarded through the network, we randomly choose another instance $I_{w,p_2}$ with the same word label but different speaker label in the training set. The word embeddings $e_{w,p_1},e_{w,p_2}\in\mathbb{R}^d$ encoded by the extractor E are
\begin{equation}
\begin{aligned}
    e_{w,p_1}=E(I_{w,p_1})\\e_{w,p_2}=E(I_{w,p_2}),
\end{aligned}
\end{equation}
where $d$ denotes the dimension of the word embeddings.
The loss function is used to calculate the distance between $e_{w,p_1}$ and $e_{w,p_2}$. 
In this paper, we investigate mean square error (MSE) regression loss as loss funtion 
\begin{equation}
    l_{MSE}(e_{w,p_1},e_{w,p_2})=\frac{1}{d} \|e_{w,p_1}-e_{w,p_2}\|^2_2,
\end{equation}
which calculates the average square difference between two embeddings of the same word spoken by different speakers. $\| \cdot \|_2$ denotes the $L_2$ norm. The variability-invariant loss and the word discrimination loss, which is typically a cross-entropy loss, jointly train the acoustic word embedding network and make the system achieve more robustness. The total loss function $L_t$ can be represented as 
\begin{align*}
    L_t=l_{CE}(y_{w,p_1},\hat{y}_{w,p_1})+&l_{CE}(y_{w,p_2},\hat{y}_{w,p_2})\\+&\alpha l_{MSE}(e_{w,p_1},e_{w,p_2}).
\end{align*}
$\hat{y}$ denotes the logit output and $y$ represents the ground truth value. The hyper-parameters $\alpha$ of the network are fixed according to the results of our pre-experiment.

\subsection{Template matching}

We employ cosine distance computation with a sliding window as our template matching scheme.

 \textsl{a) Sliding window.} We apply a fixed size sliding window over the time axis of the search content to convert an utterance into a segment sequence denoted as $\bm{y}_1, \bm{y}_2, \bm{y}_3, ..., \bm{y}_T$. Each segment is fed into the trained deep CNN, and we can get a sequence of acoustic word embedding $f(\bm{y}) = (f(\bm{y}_1),f(\bm{y}_2),...,f(\bm{y}_T)) $ derived from the output of GAP layer. Then, we pad or clip the keyword audio to make sure the length of segments equal to the size of the sliding window. Next, the  input segment $\bm{x}$ is transformed into the embedding $f(\bm{x})$ with deep CNN. So far, we can compute the cost between a segment sequence of the search content $\bm{y}$ and a spoken query $\bm{x}$ as follows:
\begin{equation}\label{equ:cost}
    Cost(\bm{x}, \bm{y})=\min{(1-\frac{f(\bm{x})\dot f(\bm{y}_i)}{||f(\bm{x})||_2||f(\bm{y}_i)||_2})}, i=1,...,T.
\end{equation}
% When the cost is significant, it means that the audio segments belong to different keywords, and when the cost is small, it means that the two segments have the same keyword.

\textsl{b) Simple moving average.} After the computation of cosine distance with the sliding window, a time-dependent score sequence is generated for each utterance to be searched. To reduce the influence of random fluctuation of the scores, we further employ simple moving average (SMA) to smooth the sequence. In SMA, output scores are calculated by taking the sum of recent scores and then dividing that by the number of involving frames for each point.

\textsl{c) Multi-template.} We also employ the multi-template strategy in our original QbE system for the same reason stated in the baseline system but with a different fusion strategy. Compared with the input features with arbitrary length in the baseline system, features extracted by the AWE system are fixed-dimensional, which allows us to directly calculate a fused embedding by computing the average values over all multiple templates. 

% There are two significant advantages of our AWE-QbE system fusion strategy over the baseline system. First, we do not have prior knowledge of how to choose a representative language-independent BNF template in the baseline system. And in the AWE system, simply averaging over templates treats each template equally and brings more robustness. Second, it's evident that simply averaging has less time consumption than DTW alignment. 

\section{Experiments}

\subsection{Experiment setup}

% \begin{figure*}[t]
%     \centering
%     \subfigure[MAP]{
%     \begin{minipage}[t]{0.33\linewidth}
%     \centering
%     \includegraphics[scale=0.4]{fig/map1.jpg}
%     \end{minipage}%
%     }%
%     \subfigure[P@5]{
%     \begin{minipage}[t]{0.33\linewidth}
%     \centering
%     \includegraphics[scale=0.4]{fig/pat51.jpg}
%     \end{minipage}%
%     }%
%     \subfigure[P@N]{
%     \begin{minipage}[t]{0.33\linewidth}
%     \centering
%     \includegraphics[scale=0.4]{fig/patn1.jpg}
%     \end{minipage}%
%     }%
%     \caption{Performance of AWE, language independent BNF+S-DTW and PPP+S-DTW systems with different number of templates.}
%     \label{fig:multitemplate}
%     \vspace{-0.4cm}
% \end{figure*}

In the PPP + S-DTW baseline system, we trained three acoustic models with the following training set: a) Chinese data, b) English data as the first language (L1), and c) mixed English data of L1 and L2 (second language). The Chinese acoustic model (a) is trained with MFCC-pitch features from the AISHELL-2 Chinese dataset \cite{du2018aishell}. English acoustic model (b) is trained with MFCC features from Librispeech dataset\cite{panayotov2015librispeech}. English L2 and L1 mixed model (c) is trained with MFCC features from the Librispeech dataset (L1 for English) and MDT-ASR-A004 dataset\footnote{https://www.magicdatatech.com/goods/3309.html} (L2 for English) . The PPP features are extracted from the output of TDNN acoustic models implemented with Kaldi nnet3 scripts \cite{povey2011kaldi}.

As for our proposed AWE system, we adopt 220k spoken word tokens for English and Chinese. Audio of 1459 English word types and 1956 Chinese word types are aligned from the Librispeech English (L1) dataset, MDT-ASR-A004 English (L2) dataset, and AISHELL-2 dataset.  
For evaluation, we select keyword audio templates from English (L1) dataset TIMIT \cite{garofolo1993darpa}, English (L2) dataset MDT-ASR-A004 and Chinese dataset THCHS30 \cite{wang2015thchs}. 86 Chinese keywords and 52 English keywords, with five to ten templates for each word, are used. The word types chosen from L1 and L2 English datasets are the same. As the L2 English keywords for evaluation and training come from the same dataset, we specially split the dataset so that no audio spoken by one person appears both in training and testing set. And we utilize 1191 utterances of code-switching dataset from Datatang AI Dataset \footnote{https://www.datatang.com} where our chosen keywords appear as the testing data. In this dataset, the speaker may alternate language from Chinese to English in some words while speaking. Our task is to detect the occurrences of both Chinese and English keywords in these code-switching audio utterances.

We employ the 64-dimensional Fbank energies as input acoustic features for our AWE system. The neural network model is trained with categorical cross-entropy and variability-invariant loss as loss function and optimized by Stochastic Gradient Descent (SGD) with Nesterov momentum 0.9. The learning rate is first initialized as 0.1 and reduces when the loss stops decreasing. We train the model for 80 epochs, and after training, we extract embeddings for segments of keyword audio and search content from the penultimate layer of the network. The size of the sliding window is 0.8 seconds, which covers the length of most keywords.

In our experiments, following previous researches \cite{chen2016unsupervised, Yuan2018Learning}, we use MAP, P@5, P@N as our evaluation metrics. MAP (Mean Average Precision) refers to the mean of average precision for each keyword in search content. P@5 (Precision at 5) is the precision of the top 5 utterances retrieved by the system and P@N (Precision at N) is the precision of top N utterances, where N means the number of target keywords in search content.

% \subsection{Multi-template}
% To explore the influence of multiple templates in the AWE system, we make a comparison between baseline system and our proposed AWE system with a different number of templates. From Fig.\ref{fig:multitemplate}, we can find that with the usage of multiple templates, performances of both baseline system and AWE system become better on MAP, P@5, and P@N. To be specific, the performance grows fast at the beginning and gradually becomes stable. We can also find that the performance of the AWE system is better than the PPP+ S-DTW systems and the gain brought by multiple templates is slightly larger than baseline system. 
% The results prove the effectiveness of the usage of multiple templates. Employing multiple templates can reduce the influence of other speech attributes, which makes the extracted embeddings more representative. To some extent, more templates bring larger improvement, while too many templates would not bring further benefit. Besides, unlike DTW fusion in baseline systems, which requires us to select a template as the main template, one of the advantages of the AWE multi-template fusion scheme is that it treats every template equally.
\begin{table}[t]
    \caption{Performance of PPP + S-DTW, one and block softmax AWE systems without variability-invariant loss and one softmax with variability-invariant loss on code switching dataset}
    \label{tab: results1}
    \centering
    \scalebox{0.85}{
    \begin{tabular}[c]{l|c|c|c|c}
        \toprule \hline
        \multirow{2}{*}{\textbf{System}} &  \multirow{2}{2cm}{\textbf{KW lang and template types}} & \multicolumn{3}{c}{\textbf{Metrics}} \\
        \cline{3-5}&& MAP & P@5 & P@N\\
        \hline\hline
        \multirow{3}{3cm}{(a) CN PPP + S-DTW} & 
        CN & 0.795 & 0.820 & 0.464 \\
        & EN (L1) & 0.046 & 0.053 & 0.036 \\
        & EN (L2) & 0.092 & 0.130 & 0.077 \\
        \hline
        \multirow{3}{3cm}{(b) EN (L1) PPP + S-DTW} & 
        CN & 0.069 & 0.113 & 0.061 \\
        & EN (L1) & 0.307 & 0.369 & 0.223 \\
        & EN (L2) & 0.284 & 0.315 & 0.212 \\
        \hline
        \multirow{3}{3cm}{(c) EN (L1,L2) PPP + S-DTW} & 
        CN & 0.144 & 0.206 & 0.113 \\
        & EN (L1) & 0.418 & 0.407 & 0.266 \\
        & EN (L2) & 0.747 & 0.726 & 0.460 \\
        \hline
        \multirow{3}{3cm}{(d) EN (L1,L2),CN   PPP + S-DTW} & 
        CN & 0.691 & 0.739 & 0.423 \\
        & EN (L1) & 0.227 & 0.284 &  0.180\\
        & EN (L2) & 0.565 & 0.642 & 0.396\\
        \hline 
        \multirow{3}{3cm}{(e) Block Softmax AWE without V-I loss} & 
        CN & 0.725 & 0.742 & 0.426 \\
        & EN (L1) & 0.556 & 0.588 & 0.377 \\
        & EN (L2) & 0.757 & 0.777 & 0.478\\
        \hline
        \multirow{3}{3cm}{(f) One Softmax AWE without V-I loss} & 
        CN & 0.701 & 0.746 & 0.422 \\
        & EN (L1) & 0.570 & 0.596 & 0.384 \\
        & EN (L2) & 0.769 & 0.596 & 0.490\\
        \hline        
        \multirow{3}{3cm}{(g) One Softmax AWE with V-I loss} & 
        CN & 0.702 & 0.737 & 0.418 \\
        & EN (L1) & 0.634 & 0.665 & 0.414 \\
        & EN (L2) & 0.804 & 0.838 & 0.534\\
        \hline\bottomrule
    \end{tabular}}
    \vspace{-0.4cm}
\end{table}

% \begin{table}[t]
%     \caption{Performance one softmax AWE without variability-invariant loss and one softmax AWE with variability-invariant loss on code switching dataset}
%     \label{tab: results2}
%     \centering
%     \scalebox{0.9}{
%     \begin{tabular}[c]{l|c|c|c|c}
%         \toprule \hline
%         \multirow{2}{*}{\textbf{System}} &  \multirow{2}{*}{\textbf{KW lang}} & \multicolumn{3}{c}{\textbf{Metrics}} \\
%         \cline{3-5}&& MAP & P@5 & P@N\\
%         \hline\hline
%         \multirow{3}{3cm}{(a) One Softmax AWE without V-I loss} & 
%         CN & 0.701 & 0.746 & 0.422 \\
%         & EN (L1) & 0.570 & 0.596 & 0.384 \\
%         & EN (L2) & 0.769 & 0.788 & 0.490\\
%         \hline 
%         \multirow{3}{3cm}{(b) One Softmax AWE with V-I loss less data} & 
%         CN & 0.702 & 0.737 & 0.418 \\
%         & EN (L1) & 0.634 & 0.665 & 0.414 \\
%         & EN (L2) & 0.804 & 0.838 & 0.534\\
%         \hline\bottomrule
%     \end{tabular}}
%     \vspace{-0.4cm}
% \end{table}
\subsection{PPP + S-DTW systems}

As Table~\ref{tab: results1} shows, system (a) and system (b) are trained with Chinese and English (L1) datasets separately, and we can see that they produce better results on their own language while lower scores on the other language. And system (a) with Chinese templates achieves the best results among all methods, which means that the PPP + S-DTW system is suitable in a single language QbE-STD task.

From the results of (b), we can observe that the scores on both L1 and L2 English keywords are lower than expectations comparing with those Chinese words. The possible explanation is that the English audio words are spoken by Chinese speakers and have different kinds of accents. Besides, the audio context of the English words is still in Chinese, which may affect the searching performance of English words. To further investigate this phenomenon, in system (c), we train an English DNN model with both L1 English and L2 English audio data to reduce the influence of accent. The results of (c) are better than (b) on both L1 and L2 English keywords, which reveals that multi-condition training can reduce the gap of audio words caused by accent. Besides, the result on Chinese keywords achieved by system (c) is better than (b), which might be because L2 English (spoken by Chinese) words contribute to Chinese word representation.

To achieve better overall results on both Chinese and English words, in the system (d), we concatenate the PPP extracted with system (a) and (c) on the feature level. The result shows that the performance of concatenated features is more balanced than the PPP of a single language.

\subsection{AWE systems with different softmax functions}

From the view of system (e) and (f), we can see that our proposed systems produce competitive results over the baseline PPP + S-DTW systems on all types of keywords except result on Chinese keywords achieved by system (a). The results on L1 English words of the proposed AWE systems are much better than other systems, which proves that our proposed method has the potential to overcome the mismatch caused by accent between keyword templates and searching content to some extent. Also, system (e) and (f) get good performance on L2 English keywords. Our proposed system is suitable for the code-switching scenario, while we should also find that on Chinese words, the performance gap between the AWE system and PPP + S-DTW system (b) still exists. The PPP + S-DTW system is still robust on a single language searching task.

\subsection{AWE systems with variability-invariant loss}
% In this work, the hyper-parameter of the variability-invariant loss is fixed at 0.8 according to our experiment results. Table \ref{tab: results2} shows the effectiveness of the usage of variability-invariant loss. We can see system (b) has better results on all types of keyword languages. The improvement of MAP on English (both L1 and L2) keywords is more considerable than Chinese keywords. It is probably because the loss is employed on instances with the same word label but different speaker labels, each English word type contains both L1 and L2 spoken token, which include more generalization information than the Chinese training dataset. 

In this work, the hyper-parameter of the variability-invariant loss is fixed at 0.8, according to our preliminary experimental results. Table \ref{tab: results1} also shows the effectiveness of the usage of variability-invariant loss. We can see that system (g) has a considerably better result on English (both L1 and L2) keywords than Chinese keywords. The loss is employed on instances with the same word label but different speaker labels. In the English scenario, it minimizes the difference within a word between not only speakers but also accents.

\section{Conclusions}

In this paper, we propose an AWE QbE-STD system based on a deep convolutional neural network. We utilize training data of two languages to train deep neural networks with both one softmax and block softmax layer. Acoustic word embeddings are extracted from the penultimate layer of the network, and cosine distances are computed between embeddings of keyword audio segments and search content segments with sliding windows. Experimental results show that our proposed AWE system with one softmax or block softmax layer generates competitive results over the baseline PPP + S-DTW systems. Variability-invariant loss is employed to decrease the influence caused by speaker-related information, and the experiment result shows the effectiveness of this method.

\section{Acknowledgements}

This research is funded in part by the National Natural Science Foundation of China (61773413), Key Research and Development Program of Jiangsu Province (BE2019054), Six talent peaks project in Jiangsu Province (JY-074), Science and Technology Program of Guangzhou, China (202007030011, 201903010040).

\newpage
\bibliographystyle{IEEEtran}

\bibliography{references}

% Generated by IEEEtran.bst, version: 1.13 (2008/09/30)
\begin{thebibliography}{10}
\providecommand{\url}[1]{#1}
\csname url@samestyle\endcsname
\providecommand{\newblock}{\relax}
\providecommand{\bibinfo}[2]{#2}
\providecommand{\BIBentrySTDinterwordspacing}{\spaceskip=0pt\relax}
\providecommand{\BIBentryALTinterwordstretchfactor}{4}
\providecommand{\BIBentryALTinterwordspacing}{\spaceskip=\fontdimen2\font plus
\BIBentryALTinterwordstretchfactor\fontdimen3\font minus
  \fontdimen4\font\relax}
\providecommand{\BIBforeignlanguage}[2]{{%
\expandafter\ifx\csname l@#1\endcsname\relax
\typeout{** WARNING: IEEEtran.bst: No hyphenation pattern has been}%
\typeout{** loaded for the language `#1'. Using the pattern for}%
\typeout{** the default language instead.}%
\else
\language=\csname l@#1\endcsname
\fi
#2}}
\providecommand{\BIBdecl}{\relax}
\BIBdecl

\bibitem{Mandal2014Recent}
A.~Mandal, K.~Prasanna~Kumar, and P.~Mitra, ``Recent developments in spoken
  term detection: A survey,'' \emph{International Journal of Speech
  Technology}, vol.~17, pp. 183--198, 2014.

\bibitem{Park2008Unsupervised}
A.~S. Park and J.~R. Glass, ``Unsupervised pattern discovery in speech,''
  \emph{Transactions on Audio Speech \& Language Processing}, vol.~16, no.~1,
  pp. 186--197, 2008.

\bibitem{muller2007dynamic}
M.~M{\"u}ller, ``Dynamic time warping,'' \emph{Information retrieval for music
  and motion}, pp. 69--84, 2007.

\bibitem{Hou2014TheNQ}
J.~Hou, V.~T. Pham, C.-C. Leung, L.~Wang, H.~Xu, H.~Lv, L.~Xie, Z.-H. Fu,
  C.~Ni, X.~Xiao, H.~Chen, S.~Zhang, S.~Sun, Y.~Yuan, P.~Li, T.~L. Nwe,
  S.~Sivadas, B.~Ma, E.~S. Chng, and H.~Li, ``The {NNI} query-by-example system
  for {MediaEval} 2015,'' in \emph{MediaEval}, 2014.

\bibitem{Rodriguez2014High}
L.~J. {Rodriguez-Fuentes}, A.~{Varona}, M.~{Penagarikano}, G.~{Bordel}, and
  M.~{Diez}, ``High-performance query-by-example spoken term detection on the
  {SWS} 2013 evaluation,'' in \emph{Proc. ICASSP}, 2014, pp. 7819--7823.

\bibitem{Leung2016high}
C.-C. Leung, H.~Xu, J.~Hou, T.~Pham, H.~Lv, L.~Xie, X.~Xiao, C.~Ni, B.~Ma,
  E.~Chng, and H.~Li, ``Toward high-performance language-independent
  query-by-example spoken term detection for {MediaEval} 2015: Post-evaluation
  analysis,'' in \emph{Proc. INTERSPEECH}, 2016, pp. 3703--3707.

\bibitem{chen2016unsupervised}
H.~Chen, C.-C. Leung, L.~Xie, B.~Ma, and H.~Li, ``Unsupervised bottleneck
  features for low-resource query-by-example spoken term detection,'' in
  \emph{Proc. INTERSPEECH}, 2016, pp. 923--927.

\bibitem{Glass2009Unsupervised}
Y.~{Zhang} and J.~R. {Glass}, ``Unsupervised spoken keyword spotting via
  segmental {DTW} on gaussian posteriorgrams,'' in \emph{ASRU}, 2009, pp.
  398--403.

\bibitem{Yang2014Intrinsic}
P.~Yang, C.-C. Leung, L.~Xie, B.~Ma, and H.~Li, ``Intrinsic spectral analysis
  based on temporal context features for query-by-example spoken term
  detection,'' \emph{Proceedings of the Annual Conference of the International
  Speech Communication Association, INTERSPEECH}, pp. 1722--1726, 2014.

\bibitem{Kramer1991Nonlinear}
M.~A. Kramer, ``Nonlinear principal component analysis using autoassociative
  neural networks,'' \emph{Aiche Journal}, vol.~37, no.~2, pp. 233--243, 1991.

\bibitem{Anguera2013Memory}
X.~Anguera and M.~Ferrarons, ``Memory efficient subsequence {DTW} for
  query-by-example spoken term detection,'' in \emph{Proc. ICME}, 2013, pp.
  1--6.

\bibitem{harry2018concatenate}
H.~Wu, M.~Li, Z.~Cai, and H.~Zhong, ``Unsupervised query by example spoken term
  detection using features concatenated with self-organizing map distances,''
  in \emph{ISCSLP}, 2018.

\bibitem{Kamper2016Deep}
H.~{Kamper}, W.~{Wang}, and K.~{Livescu}, ``Deep convolutional acoustic word
  embeddings using word-pair side information,'' in \emph{Proc. ICASSP}, 2016,
  pp. 4950--4954.

\bibitem{Settle2016DiscriminativeAW}
S.~Settle and K.~Livescu, ``Discriminative acoustic word embeddings: Recurrent
  neural network-based approaches,'' \emph{Spoken Language Technology Workshop
  (SLT)}, pp. 503--510, 2016.

\bibitem{Settle2017Query}
\BIBentryALTinterwordspacing
S.~Settle, K.~Levin, H.~Kamper, and K.~Livescu, ``Query-by-example search with
  discriminative neural acoustic word embeddings,'' in \emph{Proc.
  Interspeech}, 2017, pp. 2874--2878. [Online]. Available:
  \url{http://dx.doi.org/10.21437/Interspeech.2017-1592}
\BIBentrySTDinterwordspacing

\bibitem{levin2017segmental}
K.~{Levin}, A.~{Jansen}, and B.~{Van Durme}, ``Segmental acoustic indexing for
  zero resource keyword search,'' in \emph{2015 IEEE International Conference
  on Acoustics, Speech and Signal Processing (ICASSP)}, 2015, pp. 5828--5832.

\bibitem{Yuan2018Learning}
\BIBentryALTinterwordspacing
Y.~Yuan, C.-C. Leung, L.~Xie, H.~Chen, B.~Ma, and H.~Li, ``Learning acoustic
  word embeddings with temporal context for query-by-example speech search,''
  in \emph{Proc. Interspeech}, 2018, pp. 97--101. [Online]. Available:
  \url{http://dx.doi.org/10.21437/Interspeech.2018-1010}
\BIBentrySTDinterwordspacing

\bibitem{danwei2020with-in}
D.~{Cai}, W.~{Cai}, and M.~{Li}, ``Within-sample variability-invariant loss for
  robust speaker recognition under noisy environments,'' in \emph{ICASSP 2020 -
  2020 IEEE International Conference on Acoustics, Speech and Signal Processing
  (ICASSP)}, 2020, pp. 6469--6473.

\bibitem{liang2018spoken}
D.~{Liang}, Z.~{Huang}, and Z.~C. {Lipton}, ``Learning noise-invariant
  representations for robust speech recognition,'' in \emph{2018 IEEE Spoken
  Language Technology Workshop (SLT)}, 2018, pp. 56--63.

\bibitem{Huang2019intel}
J.~Huang and T.~Bocklet, ``Intel far-field speaker recognition system for
  voices challenge 2019,'' 09 2019, pp. 2473--2477.

\bibitem{Yuan2019Verifying}
Y.~Yuan, Z.~Lv, S.~Huang, and L.~Xie, ``Verifying deep keyword spotting
  detection with acoustic word embeddings,'' 2019.

\bibitem{povey2011kaldi}
D.~Povey, A.~Ghoshal, G.~Boulianne, L.~Burget, O.~Glembek, N.~Goel,
  M.~Hannemann, P.~Motlicek, Y.~Qian, P.~Schwarz \emph{et~al.}, ``The kaldi
  speech recognition toolkit,'' in \emph{IEEE 2011 workshop on automatic speech
  recognition and understanding}, no. CONF.\hskip 1em plus 0.5em minus
  0.4em\relax IEEE Signal Processing Society, 2011.

\bibitem{chen2015QbE}
G.~{Chen}, C.~{Parada}, and T.~N. {Sainath}, ``Query-by-example keyword
  spotting using long short-term memory networks,'' in \emph{2015 IEEE
  International Conference on Acoustics, Speech and Signal Processing
  (ICASSP)}, 2015, pp. 5236--5240.

\bibitem{He2016Deep}
K.~He, X.~Zhang, S.~Ren, and S.~Jian, ``Deep residual learning for image
  recognition,'' in \emph{Proc. CVPR}, 2016.

\bibitem{du2018aishell}
J.~Du, X.~Na, X.~Liu, and H.~Bu, ``Aishell-2: Transforming mandarin {ASR}
  research into industrial scale,'' \emph{arXiv preprint arXiv:1808.10583},
  2018.

\bibitem{panayotov2015librispeech}
V.~Panayotov, G.~Chen, D.~Povey, and S.~Khudanpur, ``Librispeech: an {ASR}
  corpus based on public domain audio books,'' in \emph{Proc. ICASSP}, 2015,
  pp. 5206--5210.

\bibitem{garofolo1993darpa}
J.~S. Garofolo, L.~F. Lamel, W.~M. Fisher, J.~G. Fiscus, and D.~S. Pallett,
  ``{DARPA TIMIT} acoustic-phonetic continous speech corpus {CD-ROM}. {NIST}
  speech disc 1-1.1,'' \emph{NASA STI/Recon technical report n}, vol.~93, 1993.

\bibitem{wang2015thchs}
D.~Wang and X.~Zhang, ``Thchs-30: A free chinese speech corpus,'' \emph{arXiv
  preprint arXiv:1512.01882}, 2015.

\end{thebibliography}

% \begin{thebibliography}{9}
% \bibitem[1]{Davis80-COP}
%   S.\ B.\ Davis and P.\ Mermelstein,
%   ``Comparison of parametric representation for monosyllabic word recognition in continuously spoken sentences,''
%   \textit{IEEE Transactions on Acoustics, Speech and Signal Processing}, vol.~28, no.~4, pp.~357--366, 1980.
% \bibitem[2]{Rabiner89-ATO}
%   L.\ R.\ Rabiner,
%   ``A tutorial on hidden Markov models and selected applications in speech recognition,''
%   \textit{Proceedings of the IEEE}, vol.~77, no.~2, pp.~257-286, 1989.
% \bibitem[3]{Hastie09-TEO}
%   T.\ Hastie, R.\ Tibshirani, and J.\ Friedman,
%   \textit{The Elements of Statistical Learning -- Data Mining, Inference, and Prediction}.
%   New York: Springer, 2009.
% \bibitem[4]{YourName17-XXX}
%   F.\ Lastname1, F.\ Lastname2, and F.\ Lastname3,
%   ``Title of your INTERSPEECH 2020 publication,''
%   in \textit{Interspeech 2020 -- 20\textsuperscript{th} Annual Conference of the International Speech Communication Association, September 15-19, Graz, Austria, Proceedings, Proceedings}, 2020, pp.~100--104.
% \end{thebibliography}

\end{document}